\documentclass[12pt]{iopart}
\usepackage{iopams}
\usepackage{graphicx}  
\begin{document}

\title{Quasi-Deuterons in Light Nuclei}
\author{H. M{\"u}ther$^1$ and Praveen C. Srivastava$^2$}
\address{$^1$Institute for Theoretical Physics, University of T\"ubingen, Auf der
Morgenstelle 14, D-72076 T\"ubingen, Germany}
\address{$^2$Department  of Physics, Indian Institute of Technology Roorkee, Roorkee 247667, India}
\ead{herbert.muether@uni-tuebingen.de}
\vspace{10pt}

\begin{abstract}
The role of pairing correlations for nucleon pairs with isospin $T=1$ and $T=0$ 
is investigated for ground-states of nuclei in the mass region $12 \leq A \leq 42$. For that 
purpose the two-nucleon densities resulting from nuclear
shell-model calculations in one and two major shells are analyzed. Various tools are applied
in this analysis including the sensitivity of correlation effects on components of the $NN$ interaction.
Significant effects on the resulting energies 
are observed due to the formation of 
$T=0$ pairs. The formation of quasi-deuterons is maximal for symmetric nuclei
with $N=Z$. The formation of $T=0$ pairs is less sensitive to the density of
single-particle states close to the Fermi energy than the $T=1$ pairing. Therefore the 
correlations in $T=0$ pairs are relevant also for excitations across shell-closures. This robustness also
explains why $T=0$ pairing does not lead to such a clear evidence in comparing
energies of neighbored nuclei as the ``odd-even mass staggering'' due to the
formation of $T=1$ pairing.
\end{abstract}

%
\vspace{2pc}

%
\submitto{\JPG}
%
%
%

\section{Introduction}

The occurrence of pairing correlations between nucleons of the same isospin is
 a well established feature in the study of ground state properties of nuclei. 
The pairing term is an important ingredient of the nuclear mass formula to 
describe the odd-even mass staggering in the binding energies of 
nuclei\cite{ring}. 
Therefore it has been an obvious extension of mean-field or Hartree-Fock 
calculations of nuclei to account for the formation of proton-proton ($pp$) 
and neutron-neutron ($nn$) pairs by performing Hartree-Fock Bogoliubov (
HFB) calculations for nuclei all over the nuclear mass table 
(For example see Refs.\cite{brink2005,broglia2012,dean2003,fraundorfer14} 
and references cited there).

While the importance of $pp$ and neutron-neutron $nn$ pairing correlations is established, no clear evidence has empirically been observed for corresponding $pn$ correlations\cite{fraundorfer14,macchia1,isk2018}. At first sight this is rather astonishing since the proton-neutron interaction is more attractive than the interaction between like nucleons and leads to the only bound states of two nucleons in the deuteron channel $^3SD_1$. Therefore one should expect stronger two-nucleon correlations in the form of $pn$ pairing than $pp$ or $nn$ pairing also in nuclei. 

Indeed, this expectation is supported by the study of pairing correlations in isospin symmetric infinite nuclear matter.
BCS calculations for infinite nuclear matter\cite{baldo90,elgor90,kuckei} predict values for the pairing gap of $pp$ or $nn$ pairing in the
$^1S_0$ channel of the order of 1 to 2 MeV, which is in reasonable agreement with empirical data for $nn$ and $pp$ pairing in finite nuclei. Corresponding calculations for $pn$ pairing in the $^3SD_1$ channel yield much larger values for the pairing gap, which are of the order of 10 MeV \cite{alm90,vonder91,baldo92,takats93,baldo95,dressedpair05,rubts17,franc21}. Therefore one may expect that it should also be seen in finite nuclei, in particular in light nuclei with equal number of protons and neutrons.

Efforts have been made to determine the $pn$ pairing in nuclei by solving the corresponding HFB equations\cite{wolter71} or to extract corresponding correlations from wave-function of shell-model calculations in one major shell\cite{langanke96} or performing variational calculations in large model spaces, which account for superpositions of various symmetry-projected HFB states \cite{vampir}. Neither these theoretical studies nor the analysis of empirical data, as discussed above, provide any clear evidence for strong $pn$ pairing effects occurring in infinite nuclear matter.

It has been argued\cite{poves98,bertsch10,bertsch11} that the strong spin-orbit field in light nuclei may spoil the isoscalar spin 1 $pn$ pairing correlations.  Bertsch and Zuo demonstrated\cite{bertsch10} that a spin triplet pairing condensate may occur in large nuclei, where the spin-orbit term tends to become smaller. Their study is based on a Woods-Saxon description for the mean field of the nucleons and a phenomenological contact interaction to generate the pairing correlations.

More recent studies tried to investigate the modifications $T=1$ as well as $T=0$ pairing-correlations in a transition from nuclear matter to finite nuclei\cite{artur19}. 
These studies are based on realistic models for the $NN$ interaction and 
support the reduction of $T=0$ pairing due to the spin-orbit term as discussed 
by Bertsch and Zuo\cite{bertsch10} for open shell nuclei with $N=Z$ like 
$^{28}$Si. The studies in Ref.\cite{artur19} predict rather strong effects of 
$T=0$ pairing for the closed-shell nuclei $^{16}$O and $^{40}$Ca. 

In the present study we extend a systematic study of $T=0$ and $T=1$ pairing 
correlations by analyzing the results of configuration-mixing shell-model 
calculations for the ground-state of light nuclei. Special attention has been 
paid to the contributions to the energy resulting from the $pp$ and $nn$ 
interaction, which are compared to the contributions of the $pn$ interaction 
in pairs of nucleons with $T=0$ and $T=1$. 
Two-nucleon densities are investigated and analyzed in terms of a 
partial wave decomposition. 

Results of calculations for $sd$-shell nuclei, assuming an inert core of 
$^{16}$O and accounting for all configurations of valence nucleons in the 
$1s0d$ shell are presented in section 2. In section 3 we discuss corresponding 
results for nuclei around $^{16}$O and $^{40}$Ca considering configurations of 
nucleons in two major shells ($1p$ and $1s0d$ for nuclei with mass number $A$ 
around 16 and $1s0d$ and $1p0f$ for $A$ around 40). 
The conclusions from the present study  are summarized in the final section 4.

\section{Nuclei in the sd-shell}
Most of the nuclear structure calculations discussed in this manuscript have 
been performed using a realistic model for the $NN$ interaction, which is 
based on the One-Boson-Exchange Potential (OBEP) A defined in Ref.\cite{OBEPA}.
This interaction has been chosen as we also want to study the dependence 
of $NN$ correlations on various parameters of a relativistic meson-exchange 
model. As we will demonstrate below, most of the results discussed are not 
very sensitive to the $NN$ interaction used.

The OBEP A interaction has been determined by fitting the $NN$ phase shifts 
and the data of the deuteron using the Thompson equation. In order to derive 
matrix elements of a two-nucleon interaction to be used in a shell-model 
calculation we assume oscillator wave function (using an oscillator length 
$b=1.76$ fm, which is appropriate for nuclei around $^ {16}O$ and solve the 
corresponding Bethe-Goldstone equation for a starting energy of -10 MeV. 
The single-particle energies have been determined from the corresponding 
Hartree-Fock definition for $^{16}$O, which yields  -1.13 MeV, -0.26 MeV, 
and 3.96 MeV for the $0d_{5/2}$, the $1s_{1/2}$, and the $0d_{3/2}$ shell, 
respectively. The shell-model calculation have been performed using a 
Fortran code, which has been inspired and checked by comparison with the 
package ``KShell'', developed by N. Shimizu et al.\cite{kshell}.

\begin{figure}[!htbp]
\begin{center}
\caption{(Color online) Contributions to the energy per valence nucleon $A$ of 
Ne (left panel), Mg (middle panel), 
and Si isotopes (right panel)
as a function of neutron number $N$ as derived from shell-model calculations 
for the $sd$ shell 
using the $G$-matrix evaluated for the One-Boson-Exchange potential $A$ as 
described in the text. 
Results are presented for the total energy ($E_{tot}$ black dots connected 
by solid line), 
the contributions resulting from the $pn$ interaction in pairs with 
isospin $T=0$ ($pn,T=0$,
red triangles connected by dashed line) and the contributions obtained from 
$pn$ pairs 
with $T=1$ (green squares connected by dashed line), $pp$ pairs (magenta stars 
connected by dotted line) and $nn$ pairs 
(blue diamonds connected by 
a yellow line). 
 \label{enercon1.fig}}
\vskip 1cm
\includegraphics[width=5.5in]{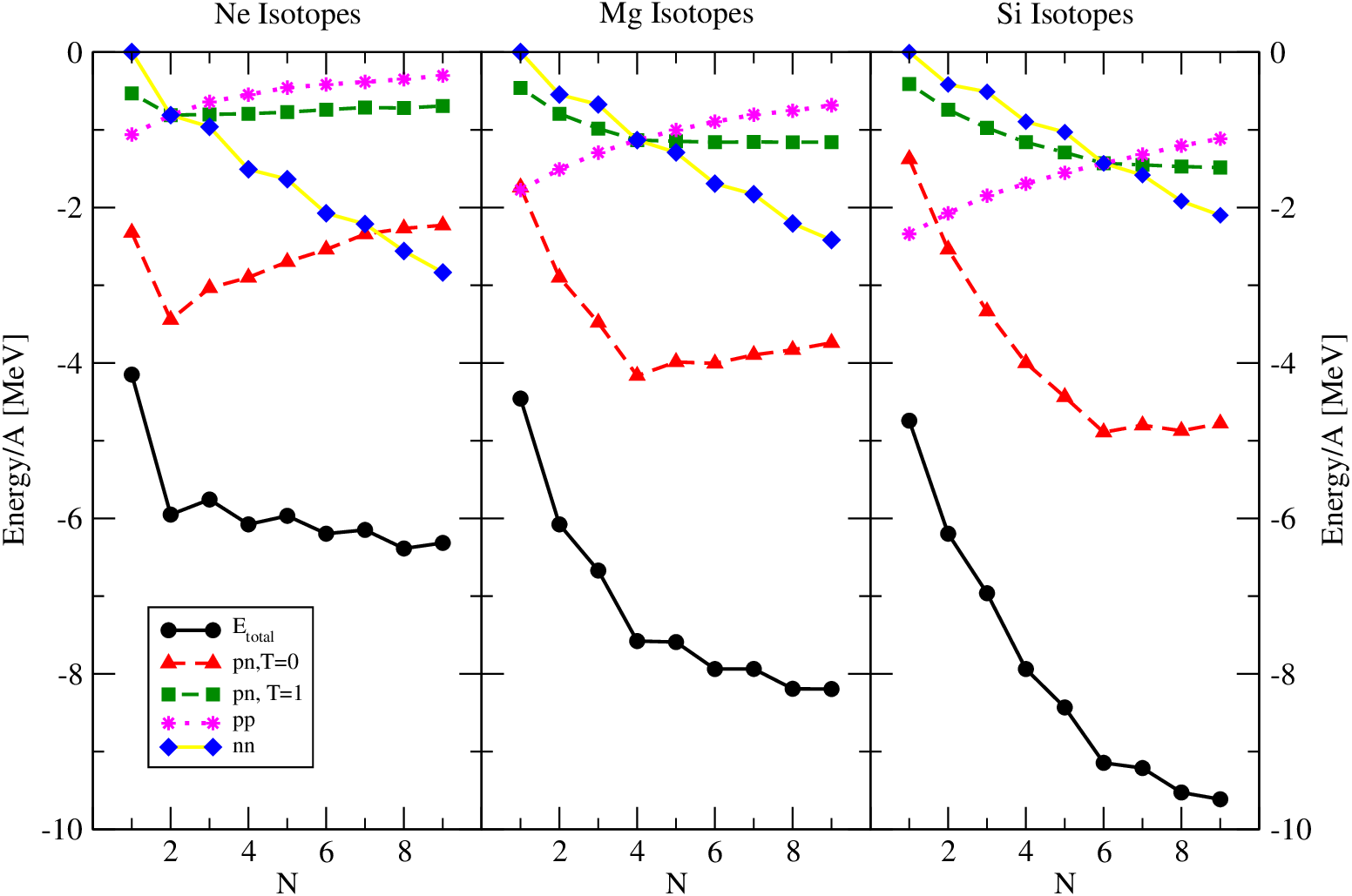}
\end{center}
\end{figure}

\begin{figure}[!htbp]
\begin{center}
\caption{(Color online) Contributions to the energy per nucleon for the 
ground-states of Ne, Mg and Si isotopes evaluated for the USD interaction of 
Brown et al.\cite{wildent}. Further details see Fig.\ref{enercon1.fig}. 
Note that for the total energy per valence nucleon a shift of 2 MeV has been 
added to allow the use of the same energy scale in the figure.  
\label{enerconw1.fig}}
\vskip 1cm
\includegraphics[width=5.5in]{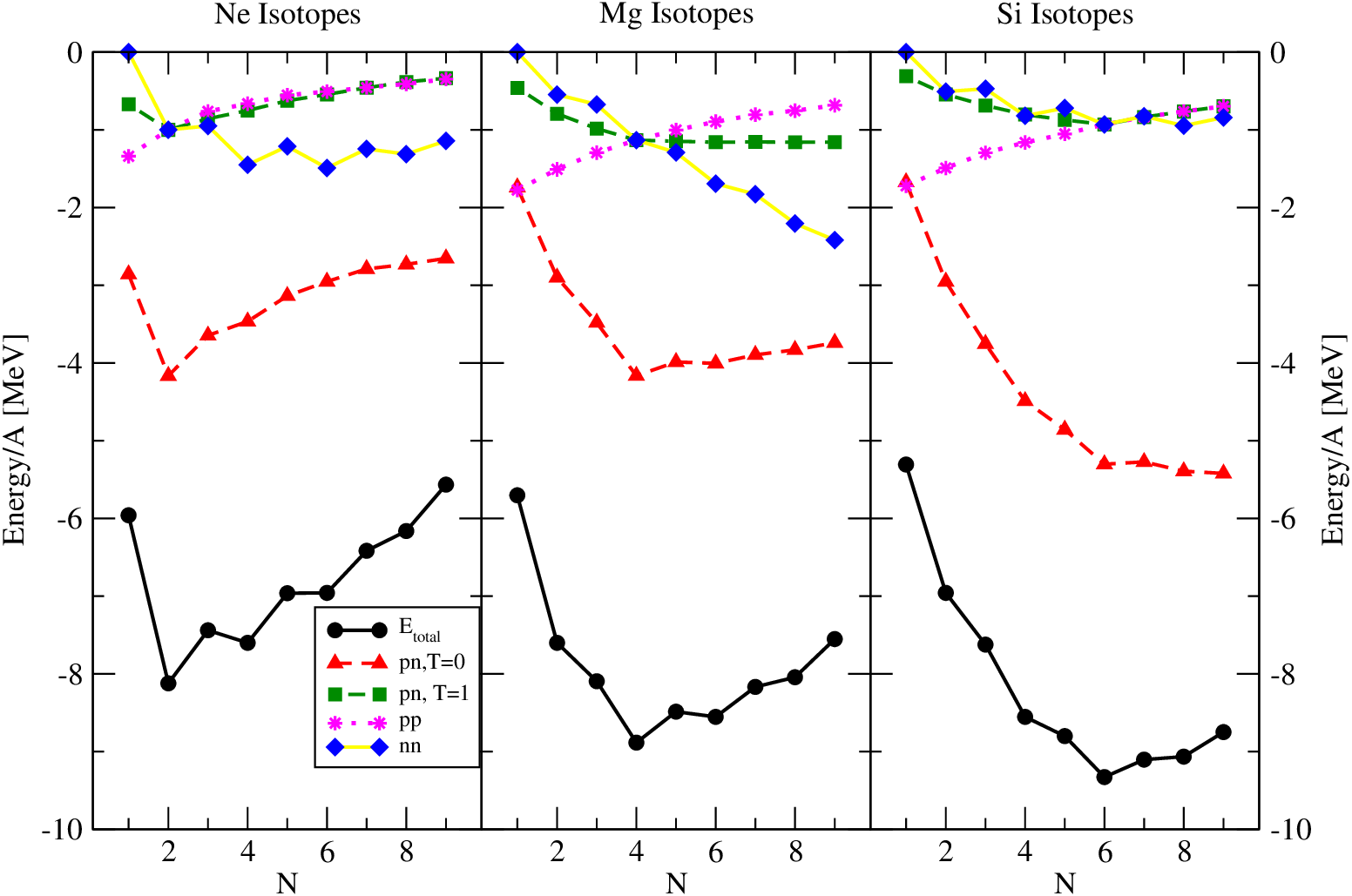}
\end{center}
\end{figure}

Results for the energy per valence nucleon and various contributions to this 
energy for the ground-states of Ne, Mg and Si isotopes are displayed in 
Fig.\ref{enercon1.fig}. Before we discuss some details for these results 
we would like to draw the attention to Fig.\ref{enerconw1.fig}, which shows the 
corresponding results, which have been obtained using quite a different 
Hamiltonian, the single-particle energies and the effective interaction USD 
defined by Brown et al. \cite{wildent}. This Hamiltonian has been fitted to 
describe nuclear data all over the sd-shell with a shell-model diagonalization.

Although the origin of the two interaction models leading to the results 
displayed in Figs.\ref{enercon1.fig} and \ref{enerconw1.fig} is rather 
different the main features of the various contributions to the energy for the 
ground states show rather similar features. Only the results for the total 
energy exhibit some differences. The fact that the binding energies obtained 
from the Brown-Wildenthal interaction are consistently larger than the 
corresponding values derived from the OBEP A interaction model is mainly due to
the fact that the single-particle energies entering the USD calculations (-3.95 MeV, -3.16 MeV, 1.65 MeV) are
more attractive as compared to the values 
(-1.13 MeV, -0.26 MeV, 3.96 MeV) used for the OBEP A calculations. 

Another difference in the calculated energies per valence nucleon is the 
feature that the Brown-Wildenthal interaction yields a minimal energy for each 
chain of isotopes displayed in Fig.\ref{enerconw1.fig} whereas the OBEP A 
interaction model leads to energies per nucleon, which tend to decrease with 
increasing neutron number. This difference is related to the fact that fit of 
the Brown-Wildenthal interaction includes a dependence of the two-body 
interaction, which weakens the interaction by  a factor 
\begin{equation}
\left(\frac{\tilde A}{18}\right) ^{(-0.3)}\,.
\end{equation}
where $\tilde A$ stands for the total number of nucleons for the nucleus 
considered. It would be possible to account for a mass dependence of the OBEP A 
interaction model by considering e.g. a mass dependence of the oscillator 
functions or reducing the starting energy in the solution of the 
Bethe-Goldstone equation. Note, however, that it is not the focus of the 
present investigation to provide an optimal model for the residual interaction. 
Instead we are interested to discuss features of the correlated many-body 
states which are general and do not depend on details of the interaction.

Such a feature is the ``odd-even mass staggering'', in the energy per nucleon 
for the various chains of isotopes displayed in Figs.\ref{enercon1.fig} 
and \ref{enerconw1.fig}. This means that nuclei with even number of neutrons 
tend to have a larger binding energy than those with odd neutron number $N$ 
can be traced back to corresponding energy contribution of neutron-neutron 
pairs, which are represented in those figures by blue triangles connected by 
yellow lines. This feature is typically interpreted as a signal for the 
formation of pairs of protons or neutrons, which is suppressed by a blocking 
effect for an odd number of protons or neutrons.

This pairing effect is a consequence of the attraction of a pair of nucleons 
with total isospin $T=1$ in the $^1S_0$ partial wave. Therefore it should be 
visible by analyzing the two-nucleon density inspecting the contribution, 
which is due to a pair of nucleons in the $^1S_0$ partial wave for the relative 
coordinates. Assuming oscillator functions, as we have done in calculating the 
interaction OBEP A described above, one can easily calculate this 
contribution using the Talmi-Moshinsky\cite{barrett} transformation and 
determine the matrix elements for the two-nucleon projection operator
\begin{eqnarray}
\langle \alpha \beta, JT \vert \hat P^{T=1} \vert \gamma \delta, JT \rangle  =  \nonumber\\
 \sum_{n,N,L} \langle \alpha\beta JT\vert n, ^1S_0, N,L,JT\rangle 
 \langle n, ^1S_0,N,L,JT\vert \gamma \delta, JT\rangle \label{eq:mosh1}
\end{eqnarray}
where $\vert \alpha\beta, JT\rangle$ denotes the antisymmetrized two-nucleon 
state for oscillator states $\alpha$ and $\beta$  coupled to total angular 
momentum $J$ and isospin $T$ whereas the quantum numbers $n$ refers to the 
radial quantum number for the relative oscillator motion in the $^1S_0$ 
partial wave while $N$ and $L$ identify the quantum numbers for the 
center of mass motion. In eq.(\ref{eq:mosh1}) 
$ \langle \alpha\beta JT\vert n, ^1S_0, N,L,JT\rangle $ denote the 
corresponding transformation brackets\cite{barrett}.

In the following we will consider the expectation value of this projection 
operator $\hat P^T_{ij}$ for the ground-state of a specific nucleus, $\langle \Psi \vert \hat P^T_{ij}\vert \Psi \rangle$  divided by the number of antisymmetrized pairs of nucleons $Q_{ij}^T$ of the kind $ij$ and isospin $T$ considered. In the case of 2 neutrons, i.e. $ij=nn$ the total isospin is $T=1$ and 
\begin{equation}
Q_{nn}^1 = \frac {N(N-1)}{2}
\end{equation} 
with $N$ the number of valence neutrons. Note that a projection operator, 
which corresponds to $\hat P^{T=1}$ in eq.\ref{eq:mosh1} can also be 
defined for the $^3S_1$ partial wave and would be applicable for proton-neutron
($pn$) pairs with isospin $T=0$ and identified as $\hat P^0_{pn}$. 
In the case, $ij=pn$,  we have for $Z$ valence protons and $N$ valence 
neutrons with $Z \leq N$ numbers of pairs with $T=0$ and $T=1$:
\begin{eqnarray}
Q_{pn}^0 & = & \frac{1}{2}Z(N+1) \nonumber\\
Q_{pn}^1 & = & \frac{1}{2}Z(N-1)\,.
\end{eqnarray}
This leads to 
\begin{equation}
\tilde S_{ij}^T = \frac{\langle \Psi \vert \hat P^T_{ij}\vert \Psi \rangle}{Q_{ij}^T}\,, \label{eq:enhanc0}
\end{equation}
The average contribution of $^1S_0$ or $^3S_1$ pairs in the two-body density of nucleon type $ij$. This quantity can be normalized dividing it by the average quantity, which is obtained when all orbits of the valence shell are completely filled
\begin{equation}
S_{ij}^T = \tilde S_{ij}^T / \tilde S_{ij}^T \mbox{(filled shell)}\,.
\label{eq:enhanc}
\end{equation} 
This will be referred to as the enhancement of the formation of $^1S_0$ or $^3S_1$ pairs in the ground-state of the nucleus under consideration.

\begin{figure}[!ht]
\begin{center}
\caption{(Color online) Enhancement of the $^1S_0$ pair contribution $S_{nn}^1$, defined in eq.\ref{eq:enhanc} for various chains of isotopes are displayed in the left panel. The right panel shows the corresponding enhancement $U_{nn}$ of the pairs with $J=0$ and $T=1$ as defined in eq.\ref{eq:enhanc0l}. \label{enhanc1s0nn.fig}}
\vskip 1cm
\includegraphics[width=3.5in]{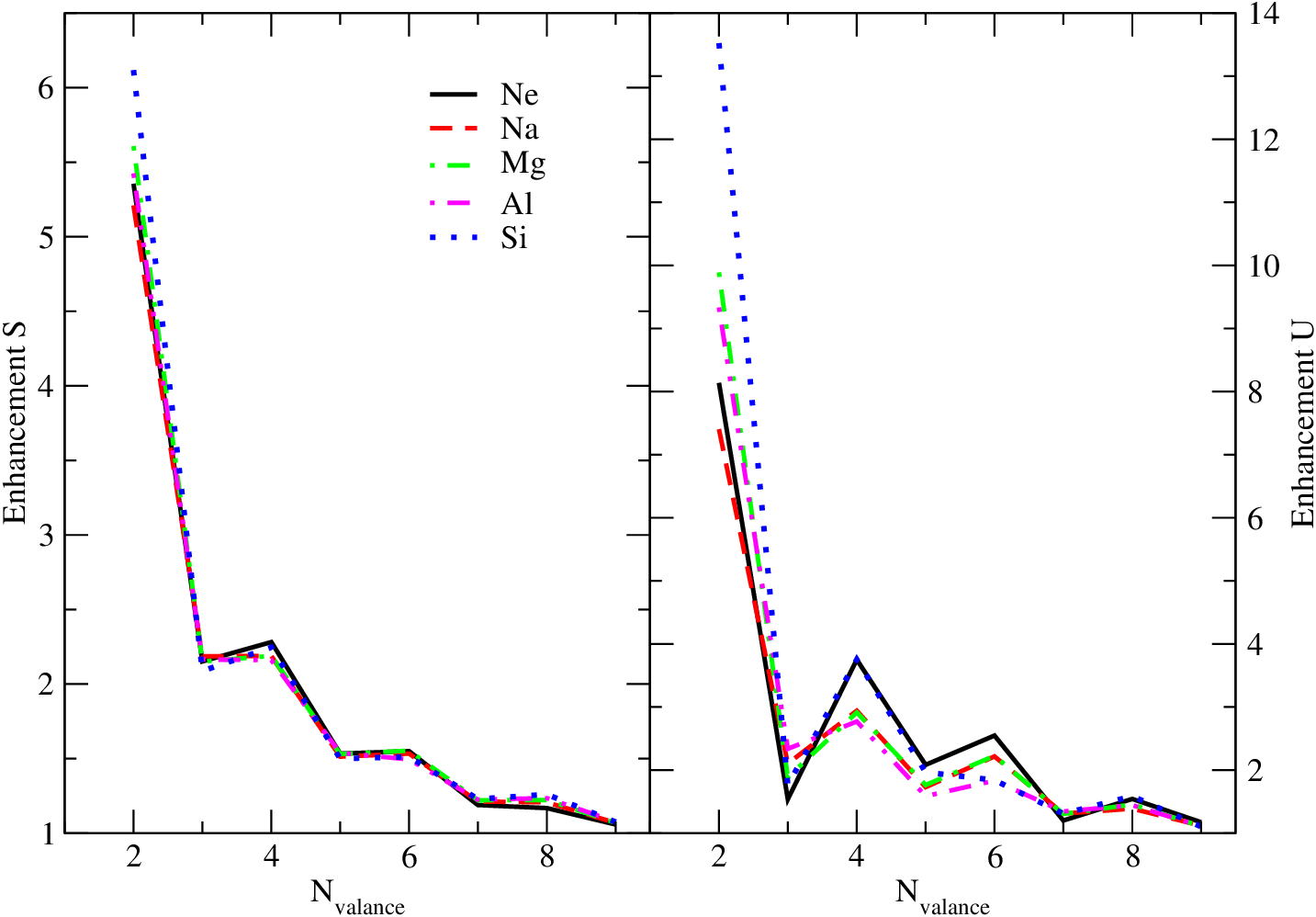}
\end{center}
\end{figure}

Results for the enhancement of $^1S_0$ neutron-neutron pairs in various 
isotopes of Ne, Na, Mg, and Si isotopes are presented in the left panel of
Fig.\ref{enhanc1s0nn.fig}. It is obvious that the enhancement factor for the 
$^{1}S_0$ component in the $nn$ part of the nuclear wave function decreases 
with the increasing number of valence neutrons $N$. For $N=2$ the components 
of the wave functions for the one pair of neutrons, which contain a large 
amplitude of the attractive $^1S_0$ partial wave are enhanced by more than a 
factor 5 as compared to the average content of  this partial wave in all 
pairs which an be formed in the $sd$ shell. For the case of e.g. $N=5$ 
one has to consider 10 orthogonal pairs of neutrons. 
Therefore the enhancement factor drops to a value around 1.5 and is bound to 
go down to $S_{nn} = 1$ for the maximal number of neutrons $N=12$ to 
fill all neutron states in the $sd$ shell. 

It is worth noting that the enhancement factor $S_{nn}^1$, which characterizes 
the $nn$ part of the two-nucleon density is almost independent on the chain 
of isotopes under consideration despite the strong interaction between 
protons and neutrons. 

Another feature, which is very evident in Fig.\ref{enhanc1s0nn.fig} is the 
``odd-even staggering'' of $S_{nn}^1$ as a function of $N$. 
This behavior  is very similar to the ``odd-even staggering'' in the binding 
energy originating from interacting neutrons displayed in 
Figs.\ref{enercon1.fig} and \ref{enerconw1.fig}.
This is a strong indication that these odd-even staggerings are related 
and indicate the formation of $nn$ pairing correlations due to the attraction 
in the $^1S_0$ partial wave in nuclei with an even number of neutrons. 

This conclusion is confirmed by inspecting the enhancement factors 
$S_{nn}^1$ as well as $S_{pp}^1$ listed in table \ref{tabel0}. 
In this table we have listed for the example of Mg isotopes enhancement 
factors resulting from the two-body densities 
obtained in shell-model calculations and compared them to those evaluated in a 
Hartree-Fock (HF) calculation. 
Note that the HF calculations are restricted to the wave functions within 
the $sd$ shell and that we consider 
the projection of the HF state $\Phi_{HF}$ to a state with good total 
angular momentum $J=0$. 
Applying the standard tools of angular momentum projection\cite{ring,ang2}, 
denoting the corresponding projection operator as $\hat P (J=0)$ one obtains
\begin{equation}
\vert \Psi_{HF} \rangle = \hat P (J=0) \vert \Phi_{HF} \rangle \,. \label{eq:angpro}
\end{equation}
The enhancement factors $S^T_{ij}$ are then obtained by replacing the SM wave 
function by this projected HF state, $\vert \Psi \rangle \to \vert \Psi_{HF} \rangle$ in eq.\ref{eq:enhanc0}.
The comparison exhibits a significant enhancement for $S_{nn}^1$ derived 
from the shell-model to the corresponding values obtained
in the mean-field approximation. This demonstrates indeed that correlations 
beyond mean-field are relevant to obtain large
components in the two-body density of $nn$ and $pp$ pair densities in the 
$^1S_0$ partial wave.

\begin{table}
\caption{\label{tabel0} 
Results for the enhancement $S_{ij}^T$ as defined in eq.\ref{eq:enhanc}  
for various isotope of Mg. Results of shell-model calculations (SM) are 
compared to those of Hartree-Fock with projection of angular momentum (HF)
}
\begin{indented}
\item[]\begin{tabular}{@{}lllll}
\br
& $S^0_{pn}$  &  $S^1_{pn}$ & $S^1_{nn}$ & $ S^1_{pp}$  \\
\mr
$^{22}$Mg  & & & & \\
SM & 1.573 & 2.182 & 5.607 & 2.282 \\
HF & 1.505 & 2.117 & 4.996 & 1.837 \\
$^{24}$Mg & & & &  \\
SM &1.529 & 2.189 & 2.189 & 2.189 \\
HF & 1.302 & 1.831 & 1.831 & 1.831 \\
$^{26}$Mg & & & & \\
SM &1.294 & 1.521 & 1.553 & 2.246 \\
HF & 1.304 & 1.528 & 1.492 & 1.776 \\
$^{28}$Mg & & & & \\
SM & 1.155 & 1.244 & 1.223 & 2.321 \\
HF & 1.159 & 1.226 & 1.131 & 1.808 \\
\br
\end{tabular}
\end{indented}
\end{table}

\begin{figure}[!ht]
\begin{center}
\caption{(Color online) Enhancement of the $^1S_0$ pair contribution $S_{pn}^1$, defined in eq.\ref{eq:enhanc}, and the corresponding enhancement factors $U_{pn}$  , defined in eq.\ref{eq:enhanc0l}, are presented for various chains of isotopes are in the left panel and right panel, respectively.  \label{enhanc1s0pn.fig}}
\vskip 1cm
\includegraphics[width=3.5in]{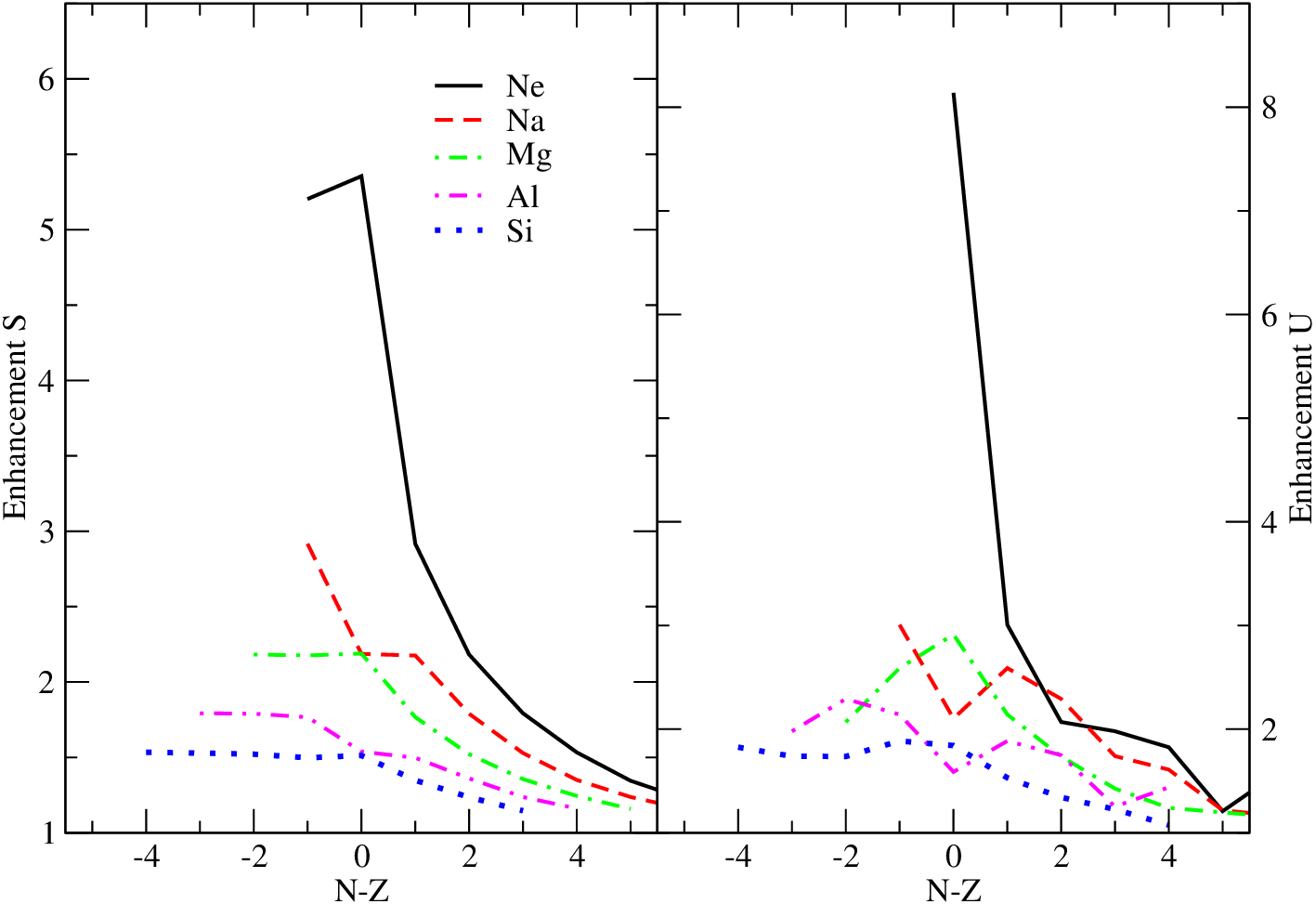}
\end{center}
\end{figure}

Results for the enhancement of the proton-neutron isovector pairing, $S_{pn}^1$, are displayed in the left panel of Fig.\ref{enhanc1s0pn.fig}.
For the nuclei with an even number of protons, i.e. $Ne$, $Mg$ and $Si$, the enhancement factor is maximal for the isotopes with $N=Z$. For the cases of an odd number of protons, $Na$ and $Al$, the enhancement $S_{pn}^1$ is suppressed for the isotopes with $N=Z$ as compared to the ones with $N=Z\pm 1$. In these cases the isovector pairing for $N=Z$, which is identical for $nn$, $pp$ and $pn$ pairs, is suppressed because the $pp$ and $nn$ pairing is suppressed for an odd number of the corresponding nucleons.

The definition of the enhancement factors $S_{ij}^{T=1}$ which we discussed so far are based
on the results of the corresponding results for the projection operator $ \hat P^T_{ij}$ defined
in eq.\ref{eq:mosh1}. This definition may be compared to the projection operator 
\begin{equation}
\mathcal{N}_{ij} = \sum_{\alpha,\beta} \vert \alpha\beta, J=0, T=1\rangle\langle \alpha\beta, J=0, T=1\vert
\label{eq:langank}
\end{equation}
counting the number of pairs of neutrons ($ij = nn$) or proton-neutron pairs ($ij=pn$) which are coupled to
angular momentum $J=0$ and $T=1$. The expectation values of this projection operator has been used
by Engels {\it et al.}\cite{engels} to investigate the role of isovector pairing in proton-rich nuclei in the pf-shell.

In analogy to eqs.\ref{eq:enhanc0} and \ref{eq:enhanc} the expectation values for the operator $\mathcal{N}_{ij}$
can be used to determine enhancement factors
\begin{equation}
 \tilde U_{ij} = \frac{\langle \Psi \vert \hat\mathcal{N}_{ij}\vert \Psi \rangle}{Q_{ij}^1}\,. \label{eq:enhanc0l}
\end{equation}
The normalized enhancement factors $U$ are defined in analogy to eq.\ref{eq:enhanc}.

Results for the corresponding enhancement factors $U_{nn}$ and $U_{pn}$ are presented in the right panels of Figs. \ref{enhanc1s0nn.fig} and \ref{enhanc1s0pn.fig}, respectively. These enhancement factors $U$, which are based on the number of nucleon pairs with $J=0$ and $T=1$ show the same features as have been discussed for the corresponding factors $S^1$ which refer to pairs of nucleons in the $^1S_0$ partial wave for the relative coordinate. One observes that the features displayed by the factors $U$ are even more pronounced than those derived from the analysis of the $S^1$ factors. 

Below, however we will focus out attention on the factors $S^T_{ij}$ as we are interested in particular in the comparison of isoscalar and isovector pairing. While $S^0_{pn}$, which is based on pairs of nucleons in the $^3S_1$ partial wave and can be interpreted to ``measure'' the amount of quasi-deuteron pairs, we do not have a corresponding definition for $U$ in the case of isoscalar pairing.

We now return to discuss the results displayed in Figs.\ref{enercon1.fig} 
and \ref{enerconw1.fig}. It is obvious from this 
that most attractive contributions to the total energy result from the 
proton-neutron ($pn$) interaction, in particular
from $pn$ pairs with isospin $T=0$. The results in these figures also indicate 
that maximal attraction from such pairs is
obtained for nuclei with $N=Z$. We want to explore, to which extent this 
maximal attraction could be identified as a signal
for an enhanced formation of quasi-deuterons or $T=0$ pairing in these nuclei. 

\begin{figure}[!ht]
\begin{center}
\caption{(Color online) Enhancement of the $^3S_1$ pair contribution $S_{pn}^0$, defined in eq.\ref{eq:enhanc} for various chains of isotopes. \label{enhanc3s1.fig}}
\vskip 1cm
\includegraphics[width=3.5in]{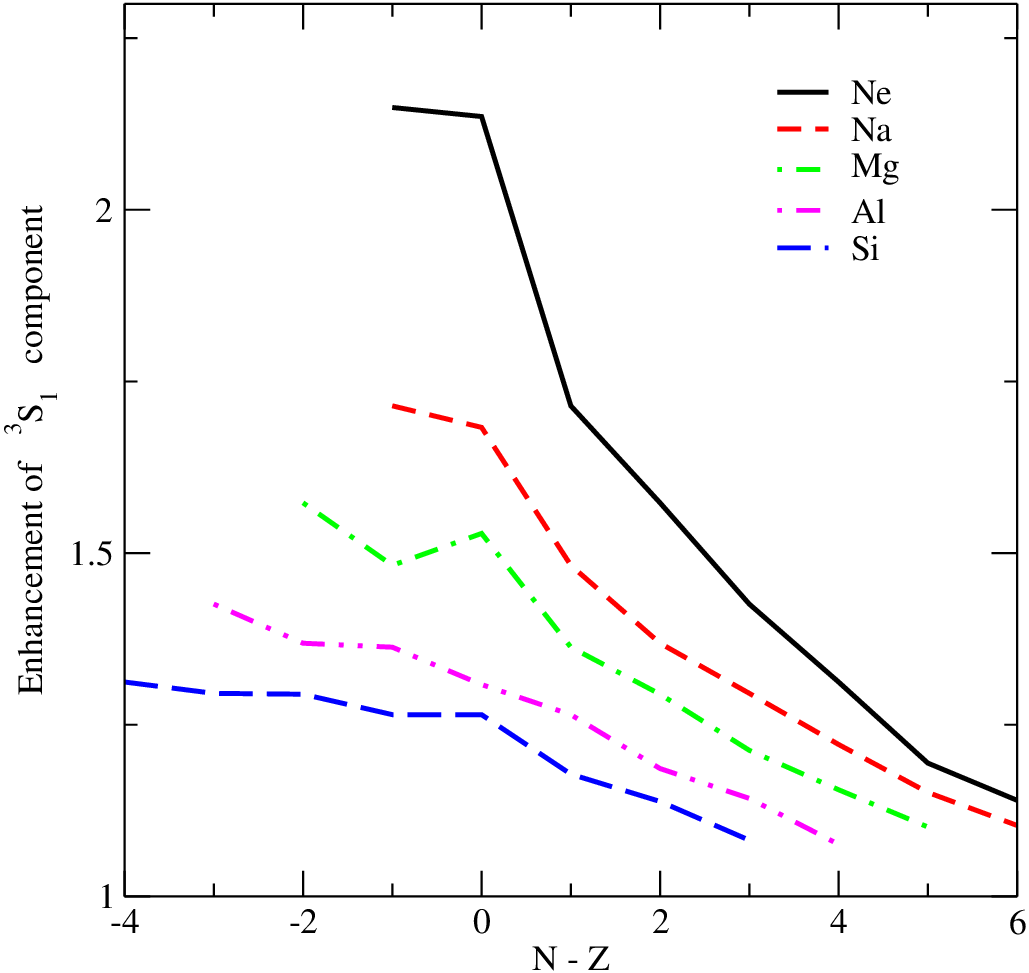}
\end{center}
\end{figure}

For that purpose we inspect the enhancement factors for the $^3S_1$ partial 
wave in the $pn$ part of the two-body densities.
This corresponds to $S_{pn}^0$ in our nomenclature defined in 
eq.\ref{eq:enhanc0}. Results 
are displayed in Fig.\ref{enhanc3s1.fig}. 

Similar to the case of $S_{nn}^1$ displayed in Fig.\ref{enhanc1s0nn.fig} also 
the values for $S_{pn}^0$ decrease 
with increasing number of valence neutrons $N$. The reason is very similar as 
discussed above: With increasing 
$N$ more $pn$ pairs compete for the attraction in the $^3S_1$ partial wave.
 
At first sight, the enhancement $S_{pn}^0$ seems to be significantly smaller 
than the results for $S_{nn}^1$.
For such a comparison, however, one should notice that e.g. for $N=2$ only 
one $nn$ pair has to be optimized and
therefore obtains a large contribution from partial wave $^1S_0$. In the 
$S_{pn}^0$ however, $N=2$ represents
3 $pn$ pairs with T=0 for the case of Ne and 4 pairs in the case of Na isotopes.
This means that the 
enhancement $S_{pn}^0$ yields similar values as $S_{nn}^1$ if we account for the number of pairs, which share
this enhancement.

A very interesting feature for our present discussion is the fact that the 
enhancement factors $S_{pn}^0$ 
plotted as a function of $N$ exhibit a kind of ``local maximum'' for the 
isotopes with $N=Z$. This means
that the maximal gain in the energy contribution from $pn$ pairs is 
accompanied by a maximum in the enhancement
of the $^3S_1$ component in the $pn$ part of the two-body density. 
Therefore and in line with the discussion
of the $T=1$ pairing between neutrons, it may be appropriate to identify this 
feature as a formation of
quasi-deuterons in $N=Z$ nuclei or as a signal of $T=0$ pairing.

This argument is supported by the comparison of enhancement factors derived 
from shell-model calculations to 
those originating from the Hartree-Fock approach listed in table \ref{tabel0}. 
We find almost no difference between
the SM and HF approach comparing $S_{pn}$ for the isotopes with $N\neq Z$, 
whereas a significant increase due to
correlations beyond mean field can be observed for the isotope $^{24}$Mg with $N=Z$.

It has been suggested by Matsubara et al.\cite{matsuba} that a careful 
analysis of the Squared Nuclear Matrix Elements (SNME) for isoscalar
and isovector M1 transitions should provide direct information for the expectation value for scalar product of the total spin of all protons $\vec\sigma_p$
and of all neutrons $\vec\sigma_n$
\begin{equation}
\langle\Psi\vert \vec\sigma_p \cdot \vec\sigma_n \vert \Psi \rangle\,. \label{eq:sigmapn}
\end{equation}
Since a pair of protons and neutrons in the isovector $^1S_0$ partial wave provides a contribution of -3/4 whereas a pair in the isoscalar partial wave $^3S_1$
 yields +1/4 the expectation value in eq.\ref{eq:sigmapn} can be interpreted as
 a measure for the relative importance of isoscalar and isovector correlations 
between protons and neutrons\cite{isacker}.

\begin{figure}[!ht]
\begin{center}
\caption{(Color online) Expectation value of the scalar product of proton spin 
and neuron spin in eq.\ref{eq:sigmapn} calculated for the ground states of 
various chains of isotopes. \label{sdpairs.fig}}
\vskip 1cm
\includegraphics[width=3.5in]{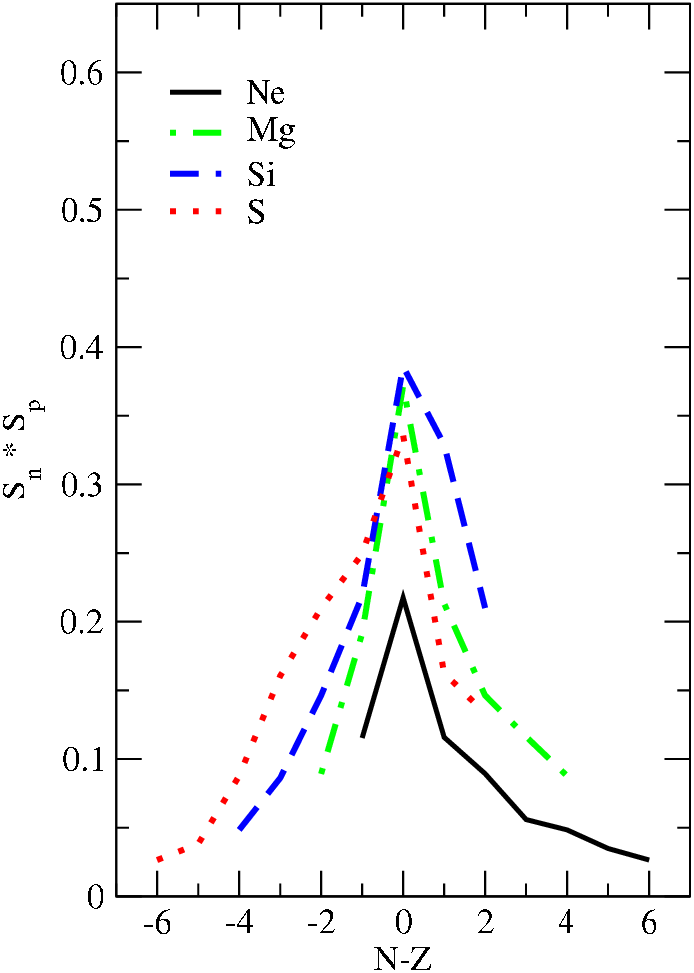}
\end{center}
\end{figure}

Results for the expectation value of the spin operator of eq.\ref{eq:sigmapn} 
are displayed in Fig.\ref{sdpairs.fig}. For all chain of isotopes considered a 
maximum is obtained for the symmetric case with $N=Z$. This suggests that the 
formation of $T=0$ pairs, or quasi-deuterons is maximal in the symmetric case. 
Note, however, that also the $^3P_J$ partial waves with isospin $T=1$ yield 
positive contributions to the product of the proton and neutron spins.

In the last part of this subsection we would like to discuss the effect of the 
ingredients of the residual interaction on the various contributions to energy
 of the ground state of nuclei in the sd-shell listed in tables \ref{tabel1} 
and \ref{tabel2}. The last column of these tables shows the excitation energy 
of the 
first excited state. For the isotopes with an even number for protons and 
neutrons, which are considered in this tables, this first excited states are 
states with 
$J=2$, while the ground-states have $J=0$. 

For each isotope the first 2 lines with label ``SM'' and ``HF''  show results 
obtained for the G-matrix derived from OBEP A using the diagonalization of the 
complete shell-model Hamiltonian and the corresponding Hartree-Fock 
approximation with angular momentum projection. One finds that all excitation 
energies
for the excite states obtained in the HF approach are significantly smaller 
than those extracted from the SM calculation. This feature may be understood
as a consequence of the formation of $T=1$ pairing correlations in the 
ground-state. At least one of these pairs must be broken to describe the 
excited state with $J=2$. Consequently, the SM approach accounting for 
pairing correlations yield a larger excitation energy than the HF 
approximation. Another 
interpretation has been given using the rotational picture for the yrast band. 
In this picture the inclusion of pairing correlation leads  to a moment 
of inertia of a superfluid, which is smaller than the corresponding moment 
of inertia for a normal liquid. Independent of the underlying interpretation: 
The increase of the excitation energy by inclusion of correlations going from 
HF to SM approximation is generally assigned to the formation of $T=1$ 
pairing correlations in the SM approach.

Other results listed in these tables in the lines with labels ``SM1'', ``SM2'', 
and ``SM3'' have been taken from shell-model calculation using a OBEP interaction
in which the contributions due to the exchange of the $\pi$-meson (SM1), 
the scalar-isoscalar $\sigma$-meson (SM2), and the $\omega$ meson has been reduced
by 10 percent. This reduction has been achieved by the reduction of the 
corresponding meson-nucleon coupling constants in the bare interaction. The
Bethe-Goldstone equation has been solved for these modified interaction and the calculation of oscillator matrix elements has been done as described above.

From the results displayed  in tables \ref{tabel1} and \ref{tabel2} one can see that a reduction of the $\pi$-exchange contribution leads to a reduction of total
binding energy, which is mainly due to the reduced attraction of $pn$ pairs with $T=0$. This can be understood from the fact that most of the attractive 
contributions to the energy of nuclei arises from the iteration of the 
strong tensor component in the coupled channels of the partial waves 
$^3S_1 - ^3D_1$.
The contributions of the NN interaction for pairs of nucleons with isospin $T=1$ are not very much affected by the reduction of the $\pi$ exchange.

It is worth noting, that the reduction of the $\pi$-exchange has a rather small
effect on the excitation energy of the first $2^+$ state. This would be in
line with the conclusion that this excitation energy is mainly affected from 
the pairing effects in the pairs of nucleons $T=1$ (see discussion above).

The reduction of the $\sigma$-exchange leads to the reduction of the attractive
components resulting from the interaction of $T=0$ pairs as well as pairs with
$T=1$. This is accompanied by a decrease of the excitation energy.

Also the reduction of the $\omega$-exchange exhibits significant effects on the 
energy contributions of all kinds of NN pairs. Since the exchange of the vector
meson $\omega$ is responsible for the short-range repulsion of the NN 
interaction, its reduction leads to an increase in the binding energy for all 
contributions $\Delta E$ as well as for the excitation energy $\Delta E_{J=2}$.

\begin{table}
\caption{\label{tabel1} 
Comparison of energy contributions to the total energy obtained in shell-model 
or Hartree-Fock calculations for 
various Neon isotopes in the sd-shell. Results are listed for the total 
energy ($E_{total}$) the contributions 
resulting from $T=0$ pairs and those from the interaction of $pn$, $pp$, 
and $nn$ pairs with isospin $T=1$. The last
column shows the excitation energy of the first excited state ($J=2$). 
The G-matrix derived from the original
OBEP A interaction yield the results presented in the lines denoted with SM 
for shell-model calculations and HF for 
corresponding Hartree-Fock calculations. The lines denoted with SM1, SM2, and 
SM3 present the results of shell-model
calculations using the OBEP A with reduced $\pi$-exchange, $\sigma$-exchange 
and $\omega$-exchange, respectively. All energies are listed in MeV.
}
\begin{indented}
\item[]\begin{tabular}{@{}lllllll}
\br
& \ $E_{total}$  &  $\Delta E_{T=0}$ &$ \Delta E_{T=1,pn}$ & $ \Delta E_{pp}$ & $\Delta E_{nn}$ &\ $\Delta E_{J=2}$ \  \\
\mr
$^{20}$Ne&\ &&&&\ & \ \\
SM &\ -23.81 & -13.78 & -3.25 & -3.25 & -3.25 \ & 1.029 \ \\
HF &\ -23.28 & -12.69 & -3.10 & -3.10 & -3.10\  &  0.779 \\
SM1 &\ -22.75 & -12.33 & -3.31 & -3.31 & -3.31 \ & 0.984 \ \\
SM2 &\ -20.65 & -12.05 & -2.70 & -2.70 & -2.70 \ & 0.755 \ \\
SM3 &\ -26.20 & -15.13 & -3.64 & -3.64 & -3.64 \ & 1.264 \ \\
\mr 
$^{22}$Ne&\ &&&& \ &\ \\
SM  &\ -36.46 & -17.40 & -4.76 & -3.28 & -9.05  \ & 0.737 \ \\
HF & \ -35.06 & -16.47 & -4.73 & -3.06 & -7.86  \ &  0.394\\
SM1  &\ -35.16 & -15.67 & -4.78 & -3.35 & -9.17  \ & 0.749 \ \\
SM2  &\ -31.62 & -15.36 & -4.01 & -2.68 & -7.30  \ & 0.510 \  \\
SM3 &\ -40.08 & -18.97 & -5.30 & -3.73 & -10.33 \ & 0.933 \ \\
\mr
$^{24}$Ne&\ &&&& \ &\\
SM  &\ -49.58 & -20.31 & -5.93 & -3.34 & -16.59  \ & 1.012 \ \\
HF & \ -47.70 & -20.44 & -6.00 & -2.83 & -14.38  \ &  0.639 \ \\
SM1  &\ -48.07 & -18.31 & -5.93 & -3.44 & -16.76  \ & 1.048 \ \\
SM2  &\ -42.81 & -17.87  & -4.96 & -2.68 & -13.45  \ & 0.735 \ \\
SM3 &\ -54.61 & -22.19 & -6.62 & -3.83 & -18.90 \ & 0.755 \ \\ 
\br
\end{tabular}
\end{indented}

\end{table}

\begin{table}
\caption{\label{tabel2} 
Comparison of energy contributions to the total energy obtained in shell-model 
or Hartree-Fock calculations 
for various isotopes of Mg in the sd-shell. Further details see caption of 
table \ref{tabel1}
}
\begin{indented}
\item[]\begin{tabular}{@{}lllllll}
\br
& \ $E_{total}$  &  $\Delta E_{T=0}$ &$ \Delta E_{T=1,pn}$ & $ \Delta E_{pp}$ & $\Delta E_{nn}$ &\ $\Delta E_{J=2}$ \  \\
\mr
$^{24}$Mg&\ &&&& \ &\\
SM  &\ -60.62 & -33.31 & -9.06 & -9.06 & -9.06 \ & 0.951 \ \\
HF & \ -54.88 & -27.35 & -7.76 & -7.76 & -7.76  \ & 0.371 \\
SM1  &\ -57.99 & -29.88 & -9.16 & -9.16 & -9.16 \ & 0.918 \ \\
SM2  &\ -52.26 & -29.09 & -7.50 & -7.50 & -7.50 \ & 0.703 \ \\
SM3 &\ -66.87 & -36.59 & -10.18 & -10.18 & -10.18 \ & 1.16 \ \\
\mr
$^{26}$Mg&\ &&&& \ &\\
SM  &\ -79.39 & -40.01 & -11.59 & -8.95 & -16.93  \ & 1.005  \ \\
HF & \ -77.01 & -40.35 & -11.59 & -7.48 & -16.30  \ & 0.580 \ \\
SM1 &\ -76.19 & -36.07 & -11.61 & -9.09 & -17.05  \ & 1.013  \ \\
SM2  &\ -68.48 & -35.04 & -9.65 & -7.29 & -13.86 \ & 0.771 \ \\
SM3 &\ -87.50 & -43.80 & -12.98 & -10.16 & -19.16 \ & 1.189 \ \\
 \mr
$^{28}$Mg&\ &&&& \ &\\
SM  &\ -98.31 & -45.98 & -13.92 & -9.06 & -26.46  \ & 1.144 \ \\
HF & \ -95.59 & -45.52 & -13.61 & -7.46 & -24.95  \ &  0.748 \ \\
SM1  &\ -94.61 & -41.52 & -13.89 & -9.21 & -26.50  \ & 1.174 \ \\
SM2  &\ -84.67 & -40.18 & -11.55 &  -7.36 & -21.62  \ & 0.942 \ \\
SM3 &\ -108.43 & -50.40 & -15.62 & -10.31 & -29.99 \ & 1.305 \ \\
\mr
\end{tabular}
\end{indented}

\end{table}

\section{Correlations beyond one major shell}

While the preceeding section has been devoted to the role of $NN$ correlations 
within one major shell, the $1s0d$ shell, we would like to add some remarks
about the importance of $NN$ correlations due to excitations across major 
shells. For that purpose we have performed shell-model calculations assuming a
core of $^4He$ and considering configurations of nucleons in the $0p$ and 
$1s0d$ shell. For the residual interaction we considered matrix elements of 
the $G$-matrix for the OBEP A interaction discussed before calculated in a 
basis of oscillator states with an oscillator length $b=1.76$ fm, which is 
appropriate for nuclei around $^{16}$O. The single-particle term has been 
calculated from the interaction with the nucleons in the core of $^4$He. 
Effects of spurious center of mass motions have been considered by 
considering an oscillator field for the center of mass motion and subtracting 
the resulting energy for the c.m. motion.

\begin{table}
\caption{\label{tabel3} 
Comparison of energies for shell-model calculations in the $0p$ + $1s0d$ shells
(upper part of the table). Listed is the gain of energy due to the admixture 
of $2\hbar\omega$ excitations
for the total energy, $ \Delta E_{total}$, 
 the contributions from interaction between pairs of nucleons wit $T=0$ and 
$T=1$ and the energy difference due to the single-particle term,
$\Delta E_{sing}$ for various nuclei around $^{16}O$ . The lower part of the 
table shows corresponding results for nuclei around $^{40}$Ca obtained in 
shell-model calculations in the $1s0d$ + $1p0f$ shells. All energies are given in MeV.
}
\begin{indented}
\item[]\begin{tabular}{@{}lllll}
\br
& \ $\Delta E_{total}$  &  $\Delta E_{T=0}$ &$ \Delta E_{T=1}$ & $ \Delta E_{sing}$   \\
\mr
$^{12}$C & -3.71 & -5.07 & -0.85 & 2.21 \\
$^{14}$N & -4.82 & -6.83 & -0.84 & 2.85 \\
$^{16}$O & -6.52 & -7.80 & -1.75 & 3.04 \\
$^{18}$O & -5.50 & -5.97 & -1.49 & 1.95 \\
$^{18}$F & -5.68 & -6.71 & -1.14 & 2.16 \\
$^{19}$O & -4.86 & -4.85 & -0.22 & 0.21 \\
$^{20}$O & -4.43 & -4.63 & -1.04 & 1.23 \\
$^{20}$Ne & -4.75 & -5.61 & -1.07 & 1.93 \\
$^{21}$Ne & -4.11 & -4.64 & 0.02 & 0.51 \\
$^{22}$Ne & -3.68 & -3.96 & -0.88 & 1.16 \\
$^{23}$Ne & -3.22 & -3.60 & -0.89 & 1.28 \\
$^{24}$Ne & -2.91 & -2.89 & -0.76 & 0.74 \\
\mr
$^{36}$Ar & -7.11 & -8.62 & -2.65 & 4.16 \\
$^{38}$Ar & -7.65 & -8.48 & -3.10 & 3.93 \\
$^{40}$Ca & -8.93 & -9.85 & -3.37 & 4.29 \\
$^{42}$Ca & -8.30 & -9.09 & -3.19 & 3.98 \\
$^{42}$Sc & -8.40 & -9.55 & -3.06 & 4.21 \\
\br
\end{tabular}
\end{indented}

\end{table}

For the various nuclei listed in the upper part of table \ref{tabel3} we have performed shell-model calculations with two different kinds of truncation schemes. 
In the first one, we considered configurations with minimal, i.e. 
0 $\hbar\omega$ excitations in the oscillator model for the single-particle 
configurations. This means for the nuclei $^{12}$C, $^{14}$N, and $^{16}$O we 
suppress all configurations with nucleons occupying states in $1s0d$ shell and 
consider a closed 
core of $^{16}$O for the nuclei with $A > 16$. In the second truncation scheme 
we consider all configurations with up to 2 $\hbar\omega$ excitations. 
This means for  the nuclei $^{12}$C, $^{14}$N, and $^{16}$O we consider 
configurations with up to 2 nucleons in states of the $1s0d$ shell and for 
the nuclei
with $A > 6$ we allow up to 2 holes in the core of $^{16}$O.

Listed in table \ref{tabel3} is the gain in the energy of the ground state 
due to the admixture of 2 $\hbar\omega$ excitations ($\Delta E_{total}$).
Also given are the contributions to this energy
difference, which are due to the interaction of $pn$ pairs with isospin 
$T=0$ ($\Delta E_{T=0}$), the interaction of nucleons with $T=1$ 
($\Delta E_{T=1}$),
and  the single-particle term,
$\Delta E_{sing}$.

\begin{figure}[!ht]
\begin{center}
\caption{(Color online) Gain in energy due to the admixture of 2 $\hbar\omega$ configurations for nuclei
around $^{16}O$.  \label{Fig7.fig}}
\vskip 1cm
\includegraphics[width=3.5in]{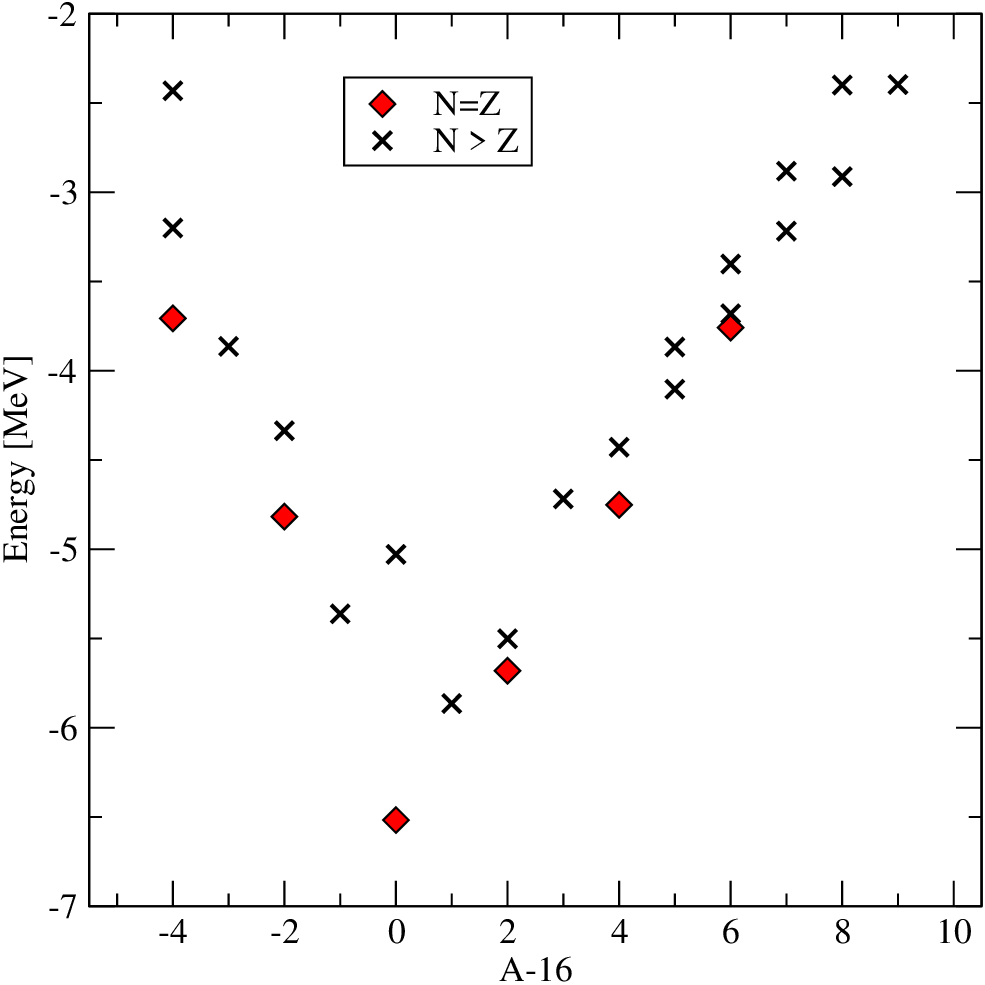}
\end{center}
\end{figure}

The total gain in energy due to the 2 $\hbar\omega$ excitations is remarkable. 
It is as large as -6.51 MeV in the case of $^{16}O$. This gain in energy 
seems to be converging: If one considers configurations up to 4 $\hbar\omega$ 
excitation the gain in energy raises to -7.15 MeV. The gain in energy is 
largest for the 
``double-magic'' nucleus $^{16}$O and decreases if one considers nuclei with 
more or less nucleons. This is also displayed in Fig. \ref{Fig7.fig}, where the gain in energy due
to the admixture of 2 $\hbar\omega$ excitations is plotted for various isotopes as
a function of the nucleon number $A$. 

This can be interpreted as a kind of blocking effect:
The 2 $\hbar\omega$ excitations correspond to excitations two-particle 
two-hole excitations of the core nucleus $^{16}$O. These configurations are 
partially
``blocked'' when additional particle and hole-states are occupied by adding or 
subtracting nucleons with respect to $N=Z=8$.

The gain in energy due to inter-shell excitations is dominated by the $T=0$ 
contributions. This dominance of $T=0$ correlations is much larger than in the 
case
of correlations within one major shell. As an example we consider the energy 
contributions to the energy of $^{20}$Ne listed in the first line of 
table \ref{tabel1}:
The contribution resulting from interaction in nucleon pairs with $T=0$, 
-13.78 MeV, is by a factor of 1.41 larger than the energy contribution from 
$T=1$ pairs.
This may be compared to the case of inter-shell correlations with a ratio 
for $^{20}$Ne of 5.24. This dominance of $T=0$ inter-shell configurations also explains that
the correlation energies are large in particular for the isotopes with $N=Z$ as can be seen
from the results displayed in Fig. \ref{Fig7.fig}.

This is in line with the results from studies of quasi-nuclear systems 
in \cite{artur19}, which demonstrated that the formation of quasi-deuterons 
or $T=0$ pairing
is much more robust against the density of single-particle states around the 
Fermi energy. Therefore it also occurs for closed-shell nuclei. 
This robustness, however, is also the reason that it does not show such a 
clear signature as the odd-even mass difference in the case of $T=1$ pairing.

\begin{table}
\caption{\label{tabel5} 
Comparison of energies and ratio of energies as defined in the text and eq.\ref{def:deltaomega} calculated for $^{16}O$
}
\begin{indented}
\item[]\begin{tabular}{@{}lllll}
\br
& OBEPA &  SM1 & SM2 &  SM3  \\
\mr
$\langle V \rangle$ &-159.94 & -152.35 & -139.58 & -174.88 \\
$ \Delta V_{2\hbar\omega}$ & -9.56 & -8.60 & -8.57 & -10.47 \\
$\Delta \Omega$ &  & 2.18 & 0.81 & 1.02\\
\br
\end{tabular}
\end{indented}

\end{table}
Finally, we would like to explore the question, which components of the $NN$ interaction lead 
to the inter-shell correlations discussed in this subsection. For that purpose we present in table \ref{tabel5}
for the example of the nucleus $^{16}O$ the expectation value for the $NN$ interaction determined in the shell-model
calculations with the inclusion of 2$\hbar\omega$ configurations, $\langle V \rangle$ and compare this quantity
with the change in this expectation value if the 2$\hbar\omega$ configurations are suppressed
$$
\Delta V_{2\hbar\omega} = \langle V \rangle - \langle V \rangle_{0\hbar\omega}\,.
$$
These calculations have been repeated using the interactions SM1, SM2 and SM3 mentioned above, 
i.e. the interactions in which the exchange of $\pi$-, the $\sigma$, and the $\omega$ exchange are suppressed
by 10 percent. We than compare the effect of the change in the interaction in the quantity $\Delta V_{2\hbar\omega}$ 
$$
\Delta \chi_{SM1} = \frac{\left[ \Delta V_{2\hbar\omega}\right]_{OBEPA} -\left[ \Delta V_{2\hbar\omega}\right]_{SM1}}{\left[\Delta V_{2\hbar\omega}\right]_{OBEPA}}
$$
and compare this with the effect in the total energy 
$$
 \chi_{SM1} = \frac{\langle V \rangle_{OBEPA} - \langle V\rangle_{SM1}}{\langle V\rangle_{OBEPA}}\,.
$$
The ratio of these 2 quantities
\begin{equation}
\Delta \Omega = \frac{\Delta \chi}{\chi}\,,\label{def:deltaomega}
\end{equation} 
is listed in table \ref{tabel5} and shows a value of $\Delta \Omega$ = 2.18 for the case of the interaction SM1. This demonstrates that a
modification of the $\pi$-exchange contribution in the $NN$ interaction leads to an effect in the correlation energy, which is by a factor
2.18 larger than the effect in the total energy. Modifications of the $\omega$ or $\sigma$ -exchange yield values for $\Delta \Omega$ of 1.02 and 0.81, respectively.
This leads to the conclusion that the effect of the $\omega$ meson on the correlation energy is very close to its effect on the total energy, while the effect
of the $\sigma$ exchange is even weaker in the correlation energy. One should recall that in our discussion of inner shell correlations in the preceeding section we
observed a connection between the $\sigma$ exchange and isovector pairing, whereas the $\pi$-exchange had larger impact on the formation of quasi deuterons in th $T=0$ pairing. 

In table \ref{tabel5} we display results for the nucleus $^{16}O$. Results for other nuclei are different if one looks at $\langle V \rangle$ or $\Delta V_{2\hbar\omega}$, but are essentially identical for the ratio $\Delta\Omega$.

The conclusions drawn from the study of inter-shell correlations in the 
neighborhood of $^{16}$O are supported by the results for nuclei around 
$^{40}$Ca.
These results have been obtained from shell-model calculations assuming a core 
of $^{16}$O and allowing for configurations of valence nucleons in the
$1s0d$ and $1p0f$ shells. The matrix elements for the G-matrix of OBEP A have 
been calculated using an oscillator length of b = 2.06 fm.

\section{Conclusions}

The role of two-nucleon correlations for pairs of nucleons with isospin $T=0$ 
and $T=1$ has been investigated by analyzing the results of shell-model 
calculations (SM) for nuclei in the mass region $12 \leq A \leq 42$ considering 
configurations in one or two major shells. For that purpose the contributions
of the $pp$, $nn$, and $pn$ pairs with $T=0$ and $T=1$ to the energy of the 
ground states of such nuclei have been analyzed and compared to a 
decomposition of the corresponding two-nucleon densities in terms of partial
waves of relative motion. 

One finds that the gain in energy for nuclei with even number of protons or
neutrons is correlated with an enhancement of the components in the two-body
density with two nucleons in the $^1S_0$ partial wave. This is interpreted as
a strong signal for the occurrence of $T=1$ pairing for such nuclei.

This interpretation is supported by a comparison of shell-model calculations 
and Hartree-Fock calculations with angular momentum projection: The two-nucleon
densities derived from HF calculations show a smaller amount of components
in the $^1S_0$ partial wave. 

In a very similar way a strong correlation is also observed between the 
energy-contributions from $pn$ pairs with $T=0$
and the components of the $^3S_1$ partial wave in the two-body density.
Consequently, such large contributions to the energy and a corresponding 
enhancement of the $^3S_1$ in the two-body density are interpreted as a
signal for the formation of quasi-deuterons or $T=0$ pairing in those nuclei.

One finds that $T=0$ pairing is maximal for symmetric nuclei with $N=Z$ but
is important also for neighbored nuclei. The occurrence of $T=0$ pairing is
not so sensitive to the density of single-particle states at the Fermi surface
and therefore one does not observe a phenomenon comparable to the blocking
effect and the ``odd-even'' staggering of $pp$ and $nn$ pairing.

This ``robustness'' can also be seen from the fact that correlations due to
excitations across shell-closures are dominated by $T=0$ correlations, while
the effect of $T=1$ excitations are significantly smaller. This is in line
with the studies of \cite{artur19} investigating quasi-nuclear systems in the
transition from nuclear matter to finite nuclei.

The results seem to be rather insensitive to the interaction considered for the
SM calculation. Most of the results presented here are based on a $G$-matrix
calculated for a realistic One-Boson-Exchange model for the $NN$ 
interaction\cite{OBEPA}. Using the meson-exchange picture for the $NN$ 
interaction allows to explore the contributions of the various mesons to the
structure of nuclei. Such investigations show that pion-exchange has a strong 
effect on the formation of $T=0$ pairing but is almost negligible for $T=1$ 
pairing. On the other hand, the exchange of isoscalar meson ($\sigma$ and 
$\omega$ mesons) effect $T=0$ and $T=1$ pairing.

\ack
This project has been supported by the DFG grant MU 705/10-2 and by a research 
grant CRG/2019/000556 from SERB (India).

\end{document}